\documentclass[12pt]{article}

\usepackage[utf8]{inputenc}
\usepackage[T1]{fontenc}
\usepackage{lmodern}
\usepackage{microtype}

\usepackage{amsmath,amssymb,amsfonts,amsthm}
\usepackage{mathtools}
\usepackage{bm,bbm}

\usepackage{graphicx}
\usepackage{booktabs,longtable, float, array,tabularx, multirow}
\usepackage{booktabs}
\usepackage{siunitx}
\usepackage{placeins} 
\usepackage{caption}
\usepackage{subcaption}
\usepackage{hyperref}
\usepackage{xcolor,pstricks}
\usepackage{seqsplit}
\usepackage{pdflscape}
\usepackage{tikz}
\usetikzlibrary{arrows.meta, calc, positioning, shapes.geometric}
\usepackage{graphicx}
\usepackage{float}
\usepackage{caption}
\usepackage{subcaption}

\usepackage[dvipsnames]{xcolor}

\usepackage{enumitem}
\usepackage{verbatim}

\usepackage{algorithm}
\usepackage{algpseudocode}

\usepackage{tikz}
\usetikzlibrary{arrows.meta, calc, positioning}

\usepackage{comment}
\usepackage{marginnote}
\usepackage{todonotes}

\usepackage{appendix}
\usepackage{authblk}
\usepackage{ifthen}
\usepackage{xr}

\usepackage[most]{tcolorbox}
\tcbuselibrary{skins,breakable}

\newtcolorbox{stepbox}{%
  enhanced, breakable,
  colback=black!1, colframe=black, boxrule=0.6pt,
  arc=2pt, left=6pt, right=6pt, top=6pt, bottom=6pt,
  before skip=6pt, after skip=6pt,
}

\newcounter{workflowstep}

\usepackage[nameinlink,capitalise]{cleveref}
\usepackage{doi}
\usepackage[natbibapa]{apacite}
\def\spacingset#1{\renewcommand{\baselinestretch}{#1}\small\normalsize}
\spacingset{1}
\newcommand{\ds}{\displaystyle}

\newcommand{\argmin}{\operatorname*{arg\,min}}

\newcommand{\prox}{\operatorname{prox}}

\newcommand{\R}{\mathbb{R}}

\newcommand{\rank}{\operatorname{rank}}

\newcommand{\ind}{\mathbb{I}}

\newcommand{\blind}{0}

\begin{document}
\if0\blind
\title{\bf Proximal Empirical Bayes for Sparse Regression with Posterior Decision Support}
\author{Dimitrios Roxanas\thanks{d.roxanas@sheffield.ac.uk}\\
School of Mathematical and Physical Sciences, The University of Sheffield,\\ S3 7RH, United Kingdom}
\date{}
\maketitle
\fi

\if1\blind
\begin{center}
{\bf Proximal Empirical Bayes for Sparse Regression with Posterior Decision Support}
\end{center}
\fi


\begin{abstract}
In this work, we revisit estimation and variable selection in sparse Gaussian regression, possibly under affine constraints.\ We adopt a middle-ground approach between optimisation and fully hierarchical Bayesian methods, preserving much of the computational efficiency of the former while adding posterior uncertainty quantification.\ In particular, we develop an empirical Bayes framework that separates shrinkage calibration from posterior-informed selection while using the same log-concave model throughout.\ The resulting workflow is modular:\ different methods can be used for noise-variance estimation, shrinkage calibration and posterior sampling.\ At the same time, these tasks are linked, and numerical choices made at one stage can propagate through the workflow and ultimately affect the decision about which variables to retain.\ We investigate in detail one instance of the framework based on a Laplace prior, with the amount of shrinkage determined by stochastic approximation proximal gradient. The proximal machinery used for this calibration also supports posterior simulation.\ We further show how the empirical Bayes calibration should be modified when homogeneous constraints reduce the number of free parameters.\ Nonhomogeneous affine information is instead incorporated softly;\ when this information is strong, posterior sampling can become poorly conditioned, and we develop a preconditioning strategy to improve mixing and posterior exploration.\ Synthetic experiments examine how scale calibration, posterior run length and sampler conditioning affect estimation, uncertainty quantification and the final sparse decision.\ On a diabetes dataset, posterior uncertainty agrees closely with established Bayesian analyses, while the final decision provides a sparser practical summary.
\end{abstract}

\newpage

\spacingset{1.45}
\section{Introduction}\label{sec:introduction}

Sparse regression combines two tasks:\ estimating a coefficient vector and deciding which coordinates are important enough to retain.\ Penalised estimators such as the Lasso solve both through a sparse optimisation problem \citet{tibshirani1996}.\ Under a Gaussian likelihood, the same $\ell_1$ penalty corresponds to an independent Laplace prior \citet{park2008}, which also provides posterior information about coefficient magnitude, sign and uncertainty.\ Bayesian variable selection can exploit that information through spike-and-slab models, continuous shrinkage priors, posterior inclusion summaries and decision rules \citet{bondell2012,hahn2015}.\ These approaches can be highly effective, but moving from a sparse point estimate to a calibrated posterior decision also introduces additional modelling and computational choices.

Recent work has connected variable selection with optimisation and developed optimisation-based Bayesian methods for sparse and constrained problems \citet{xu2023l1ball, xu2024bayesian, zhou2024, duan2026variable}.\ A related line of work in Bayesian inverse problems shows how convex nonsmooth priors can be handled by empirical Bayes calibration, proximal optimisation and proximal Markov chain Monte Carlo (MCMC) \citet{pereyra2016, durmus2018, vidal2020part1, deBortoli2020part2, pereyra2020accelerating}.\ Moreau--Yosida smoothing is central to this construction because it replaces the nonsmooth prior potential by a differentiable approximation whose gradient is available through the proximal map.\ The resulting approximation is not tied to one sampler and can also be explored with other gradient-based methods, including Hamiltonian Monte Carlo; see, for example, \citet{zhou2024, xu2024bayesian}.

The central contribution of this paper is an end-to-end sparse-regression workflow in which calibration, point estimation, posterior approximation and the final support decision remain distinct statistical objects while sharing the same proximal structure.\ We first estimate the observation variance, $\widehat{\sigma}^2$, then calibrate a global Laplace shrinkage parameter, $\widehat\theta$, by stochastic approximation proximal gradient (SAPG), compute the nonsmoothed MAP, $\widehat{\beta}_{\rm MAP},$ by FISTA \citet{beck2009}, and use MYULA to obtain posterior uncertainty from a Moreau-smoothed approximation.\ The MAP supplies the joint sparse coefficient estimate and candidate configuration, while posterior activation probabilities determine which MAP-supported coordinates remain relevant once uncertainty is taken into account.\ The resulting pipeline is
\[
\widehat{\sigma}^2
\;\longrightarrow\;
\widehat\theta
\;\longrightarrow\;
\widehat{\beta}_{\rm MAP}
\ \hbox{and posterior samples}
\;\longrightarrow\;
\text{selected set}\; \widehat S.
\]
The separation is deliberate.\ The observation scale controls the likelihood--prior balance, the empirical Bayes step fixes the global shrinkage level, the nonsmoothed MAP remains the point estimator, and the posterior sample is used for uncertainty and the final decision rather than as a substitute for the sparse optimisation problem.\ Under a continuous shrinkage prior, posterior draws are almost surely nonzero, so selection requires a practical notion of effect size.\ We use a posterior-scale threshold and require both $|\widehat\beta_{{\rm MAP},j}|\geq\tau$ and sufficiently high posterior probability that $|\beta_j|$ exceeds the same threshold.\ A decision-level Monte Carlo diagnostic then measures whether an eligible coefficient lies close to the probability boundary relative to simulation error.

A second contribution concerns empirical Bayes calibration in the presence of affine information, extending the SAPG methodology of \citet{vidal2020part1}.\ Hard homogeneous constraints and nonhomogeneous affine relations look similar at the level of the regression coefficients but behave differently in the marginal-likelihood calibration.\ For a hard homogeneous restriction $A\beta=0$, the prior is supported on a lower-dimensional linear space and the SAPG score depends on its intrinsic dimension.\ For a nonhomogeneous relation, such as $\ds \sum_{j=1}^p \beta_j =1$, direct restriction changes the prior normalising constant in a less tractable way.\ We instead retain the ambient-space homogeneous prior and impose the relation softly through a Gaussian pseudo-observation.\ This preserves the unconstrained-homogeneous SAPG score while allowing the affine information to enter through the smooth part of the model.\ Strong soft information also creates a separate numerical issue:\ low-rank curvature can force a scalar-step Langevin sampler to move slowly in otherwise well-scaled directions.\ We therefore study a geometry-aware preconditioner that targets this stiffness.\ Both hard and soft affine structures arise in our concurrent sparse index-tracking application \citet{roxanas2025index}, where portfolio construction used a soft sum-to-one condition and rebalancing used hard self-financing changes with sum zero.

The resulting method is computationally simple.\ Laplace and elastic net-type penalties have inexpensive proximal maps, so calibration, MAP optimisation and posterior simulation can reuse the same basic numerical operations.\ Richer hierarchical shrinkage models provide greater prior flexibility, particularly across heterogeneous signal sizes, but require additional modelling, calibration and posterior computation.\ Our aim here is to determine how far a comparatively simple log-concave model can support posterior analysis and sparse decisions when its computational stages are separated and checked explicitly.

The numerical studies therefore examine the behaviour of the full workflow, not only its final recovery rates.\ Study~1 evaluates the complete unconstrained workflow in both $n>p$ and $p>n$ regimes and tracks how likelihood-scale calibration and posterior Monte Carlo precision propagate to the final sparse decision.\ Study~2 examines soft affine information, showing how the same information can improve the fitted model while creating low-rank stiffness in posterior sampling, and quantifies the effect of geometry-aware preconditioning.\ Study~3 applies the workflow to the diabetes data used in \citet{efron2004} and compares its posterior intervals with externally reported Bayesian analyses.\ The reported 95\% Bayesian zero-exclusion pattern agrees with the ProxMCMC, Bayesian lasso and horseshoe analyses reported by \citet{zhou2024}, while the MAP-plus-posterior rule gives a smaller practical support at its reference operating point.\ The supplement adds a hard-constraint calibration check and an observation-scale/Moreau sensitivity analysis.

We use the Laplace prior throughout so that the calibration and the role of the final decision remain transparent.\ The proximal optimisation and sampling components, however, extend to other proper closed convex penalties with tractable proximal maps, including elastic net penalties.\ Section~\ref{subsec: SAPG} notes what changes in the empirical Bayes calibration when the simple homogeneous Laplace score is no longer available.

The remainder of the paper is organised as follows.\ Section~\ref{sec:method} gives the model and computational methodology, Section~\ref{sec:studies} presents the numerical studies, and Section~\ref{sec:discussion} discusses the main findings and extensions.\ Further derivations and implementation details are provided in the supplement.

\section{Model and methodology}
\label{sec:method}

We work with $n$ observations and $p$ predictors:
\begin{equation}
y=X\beta+\varepsilon,\qquad
\varepsilon\sim N(0,\sigma^2 I_n),
\label{eq:model}
\end{equation}
with \(y\in\mathbb R^n\), \(X\in\mathbb R^{n\times p}\) and \(\beta\in\mathbb R^p\).\ The response and predictors are centred when an intercept is not modelled explicitly, and predictor scaling is stated for each experiment.

Conditional on \(\sigma^2\), the likelihood is 
\begin{equation}
\pi(y \mid\beta)
\propto
\exp\{-f_0(\beta)\}, \qquad f_0(\beta)=\frac{1}{2\sigma^2}\|y-X\beta\|_2^2.
\label{eq:likelihood}
\end{equation}

We use the Laplace family
\begin{equation}
\pi(\beta\mid\theta)
\propto
\exp\{-\theta g(\beta)\}, 
\qquad g(\beta) = \|\beta\|_1, \;\; \theta>0
\label{eq:prior}
\end{equation}
as the shrinkage prior.\ Conditional on $(\sigma^2,\theta)$, the posterior is
\[
\pi(\beta \mid y,\sigma^2,\theta)
\propto
\pi(y \mid \beta,\sigma^2)\,\pi(\beta\mid\theta).
\]

\subsection{Likelihood scale calibration}\label{subsec:noise}

We estimate the observation variance before the empirical Bayes step and treat the result as fixed thereafter.\ This scale matters because it controls both the likelihood--prior balance and the curvature used in the proximal calculations.

When residual degrees of freedom are available, let $q=\rank(X)<n$ and let \(\widehat\beta_{\rm LS}\) denote a least-squares solution.\ The usual Gaussian residual estimator is
\begin{equation}
  \widehat{\sigma}_{\rm res}^2
  =
  \frac{\|y-X\widehat\beta_{\rm LS}\|_2^2}{n-q},
  \qquad q=\operatorname{rank}(X)<n.
  \label{eq:resvar}
\end{equation}
If an intercept is fitted, including implicitly through sample centring, the denominator is reduced by one additional degree of freedom.

When $p\ge n$ and $X$ has full row rank, ordinary least squares leaves no residual degrees of freedom.\ Under a sparse high-dimensional model we use the scaled Lasso of \citet{sun2012} as an auxiliary estimator,
\begin{equation}
  (\widehat\beta_{\mathrm{SL}},\widehat\sigma_{\mathrm{SL}})
  \in
  \argmin_{\beta\in\R^p,\,\sigma>0}
  \left\{
    \frac{\|y-X\beta\|_2^2}{2n\sigma}
    +\frac{\sigma}{2}
    +\lambda_0\|\beta\|_1
  \right\}, \qquad \lambda_0=\sqrt{\frac{2\log p}{n}}.
  \label{eq:scaled-lasso-stage0}
\end{equation}
In finite samples, shrinkage can leave signal in the scaled Lasso residuals and inflate the resulting variance.\ We therefore use the scaled Lasso only to define the provisional support
\begin{equation}
  \widehat S_{\mathrm{SL}}
  =\{j:\widehat\beta_{\mathrm{SL},j}\neq0\},
  \label{eq:scaled-lasso-support}
\end{equation}
refit least squares on that support,
\begin{equation}
  \widehat\beta_{\mathrm{post},\widehat S_{\mathrm{SL}}}
  \in
  \argmin_{b\in\R^{|\widehat S_{\mathrm{SL}}|}}
  \|y-X_{\widehat S_{\mathrm{SL}}}b\|_2^2,
  \label{eq:postols-refit}
\end{equation}
and use
\begin{equation}
  \widehat\sigma^2_{\mathrm{postOLS}}
  =
  \frac{
    \|y-X_{\widehat S_{\mathrm{SL}}}
      \widehat\beta_{\mathrm{post},\widehat S_{\mathrm{SL}}}\|_2^2
  }{
    n-\rank(X_{\widehat S_{\mathrm{SL}}})
  }.
  \label{eq:postols-sigma}
\end{equation}
An intercept adjustment is made when needed.\ Because the support is data-dependent, \eqref{eq:postols-sigma} is used as a plug-in scale rather than as an unbiased post-selection variance estimator.\ We emphasise that the scaled Lasso and post-OLS coefficients are not used downstream.

Thus
\begin{equation}
  \widehat\sigma^2
  =
  \begin{cases}
    \widehat\sigma^2_{\mathrm{res}},
      & \rank(X)<n,\\[1mm]
    \widehat\sigma^2_{\mathrm{postOLS}},
      & \rank(X)=n\ \text{under sparse high-dimensional structure}.
  \end{cases}
  \label{eq:stage0-sigma-rule}
\end{equation}
Section~\ref{study: one} shows the finite-sample effect of this choice in the $p>n$ regime.\ Other scale estimators could be substituted at this stage, including square-root Lasso, robust residual scales and high-dimensional moment estimators under stronger assumptions.\ The supplementary sensitivity analysis separately examines how alternative supplied scales propagate through the fitted posterior and final selection rule.\ When the quadratic data-fit term is better viewed as a loss and no physical noise scale is available, the same architecture can instead use a generalised Bayes learning rate or inverse temperature.\ Our sparse index-tracking application \citet{roxanas2025index} presents such an example.

\subsection{Moreau--Yosida smoothing and MYULA}
\label{subsec:MY}

We write \(f\) for the smooth quadratic term in the negative log-posterior:\ \(f=f_0\) in the unconstrained model, and \(f=f_{\tau_c}\) when the soft affine term introduced in subsection~\ref{subsec: SAPG} is present.\ The posterior combines \(f\) with a convex but nonsmooth $\ell_1$ term.\ Proximal MCMC methods exploit this structure through Moreau--Yosida smoothing.\ For a proper closed convex function $g$, its Moreau--Yosida envelope is
\begin{equation} \label{def:prox-operator}
	g_\lambda(w) = 
	\min_{u \in \mathbb{R}^p}
	\Bigl\{ g(u) + \frac{1}{2\lambda}\|u-w\|_2^2 \Bigr\}, \qquad \lambda>0,
\end{equation}
and we write
\[
\prox_{\lambda g}(w)
=
\arg\min_{u\in\mathbb R^p}
\left\{g(u)+\frac{1}{2\lambda}\|u-w\|_2^2\right\}
\]
for the corresponding proximal map.\ The envelope has gradient
\begin{equation}
	\nabla g_\lambda(\beta)
	= \frac{1}{\lambda}\Bigl(\beta - \prox_{\lambda g}(\beta)\Bigr).
\end{equation}
For the $\ell_1$ penalty, the proximal map reduces to coordinatewise soft-thresholding.\ Moreau smoothing therefore replaces the nonsmooth term by a differentiable approximation with an explicitly available gradient;\ see, for example, \citet{durmus2018} and \citet[Section~2.1]{zhou2024}.

The smoothing parameter $\lambda$ controls both approximation and computation.\ As $\lambda\downarrow0$, the Moreau envelope converges pointwise to the original convex function and, under the conditions used for MYULA, the induced smoothed posterior converges to the nonsmoothed target in total variation \citet{durmus2018,pereyra2020accelerating}.\ Smaller values of $\lambda$ therefore sharpen the posterior approximation, but also increase the Lipschitz constant of the smoothed gradient and restrict the admissible Langevin timestep.\ The supplementary sensitivity analysis varies $\lambda$ on a problem with a synthetic dataset, allowing this approximation--efficiency trade-off to be separated from changes in the supplied observation scale.
 
Writing
 $ h_\theta(\beta)=\theta g(\beta),$
we use
\begin{equation}
  h_\theta^\lambda(\beta)
  =\min_u\left\{
  \theta g(u)+\frac{1}{2\lambda}\|u-\beta\|_2^2
  \right\}.
  \label{eq:my-envelope}
\end{equation}
Its gradient is
\begin{equation}
  \nabla h_\theta^\lambda(\beta)
  =\frac{1}{\lambda}
  \left\{\beta-\prox_{\lambda\theta g}(\beta)\right\},
  \label{eq:my-gradient}
\end{equation}
which is $1/\lambda$-Lipschitz.\ 
We define
\begin{equation}
  U_{\theta,\lambda}(\beta)
  =f(\beta)+h_\theta^\lambda(\beta),
  \qquad
  \pi_{\theta,\lambda}(\beta\mid y)
  \propto e^{-U_{\theta,\lambda}(\beta)}.
  \label{eq:smoothed-posterior}
\end{equation}

MYULA \citet{pereyra2016,durmus2018} is built upon an Euler--Maruyama discretisation of the overdamped Langevin diffusion associated with $U_{\theta,\lambda}$:
\begin{equation}\label{eq:MYULA}
	\beta^{(k+1)} = \beta^{(k)} - \delta \nabla U_{\theta,\lambda}(\beta^{(k)})
	+ \sqrt{2\delta}\,\xi^{(k+1)}, 
	\qquad \xi^{(k+1)} \overset{\text{i.i.d.}}{\sim} \mathcal{N}(0,I_p),
\end{equation}
with timestep $\delta>0$.\ In our setting, the recursion is
\begin{equation}
\begin{split}
  \beta^{(k+1)}
  ={}&\beta^{(k)}
  -\delta\left[
  \nabla f(\beta^{(k)})
  +\frac{1}{\lambda}
  \left\{\beta^{(k)}-\prox_{\lambda\theta g}(\beta^{(k)})\right\}
  \right]
  +\sqrt{2\delta}\,\xi^{(k+1)}, \quad
  \xi^{(k+1)}\overset{\text{i.i.d.}}{\sim}\mathcal N(0,I_p).
\end{split}
\label{eq:myula}
\end{equation}

The continuous-time diffusion has $\pi_{\theta,\lambda}$ as its invariant distribution under the usual regularity conditions.\ The unadjusted Euler chain introduces a further finite-step approximation, so we distinguish the nonsmoothed posterior $\pi_\theta$, the smoothed posterior $\pi_{\theta,\lambda}$ and the invariant distribution of the discretised chain, denoted $\pi_{\theta,\lambda,\delta}$.\ If $\nabla f$ is $L_f$-Lipschitz, then a global Lipschitz bound for $\nabla U_{\theta,\lambda}$ is
\begin{equation}
  L_\lambda = L_f+\frac{1}{\lambda}.
  \label{eq:Llambda}
\end{equation}

The bound
\begin{equation}
  0<\delta\leq \frac{1}{L_\lambda}
  =\frac{1}{L_f+1/\lambda}.
  \label{eq:myula-conservative-step}
\end{equation}
on the timestep ensures both numerical stability \citet{pereyra2020accelerating} and the validity of the nonasymptotic MYULA analysis of \citet{durmus2018}, and the MYULA kernels used within the SAPG methodology of \citet{vidal2020part1}.\ 
In the unconstrained Gaussian model, where \(f=f_0\),
\begin{equation}
  L_f
  =\frac{\lambda_{\max}(X^\top X)}{\widehat{\sigma}^{\,2}}.
\end{equation}
For computational efficiency, it is necessary to use values of $\delta$ that are close to the stability limit \eqref{eq:myula-conservative-step}, so we take
\begin{equation}
  \delta
  =\frac{0.9}{L_f+1/\lambda}.
  \label{eq:myula-practical-step}
\end{equation}
Our default Moreau choice is
\begin{equation}
  r_\lambda:=\lambda L_f=1,
  \qquad
  \lambda=\frac{1}{L_f},
\end{equation}
for which $L_\lambda=2L_f$ and hence
\begin{equation}
  \delta=\frac{0.9}{2L_f}.
\end{equation}

The three computational objects are consequently kept conceptually separate:
\[
\pi_\theta
\longrightarrow
\pi_{\theta,\lambda}
\longrightarrow
\pi_{\theta,\lambda,\delta},
\]
representing the nonsmooth posterior, the Moreau-smoothed posterior and the finite-step unadjusted chain, respectively.\ A Metropolis--Hastings correction could instead be added, yielding a Metropolis-Adjusted Langevin Algorithm (MALA)-type kernel that removes the finite-step discretisation bias with respect to the smoothed target \(\pi_{\theta,\lambda}\) (though not the Moreau approximation itself), at the cost of an accept--reject step; we retain MYULA here for computational efficiency.

\subsection{Empirical Bayes calibration and affine information}\label{subsec: SAPG}

With the observation scale fixed at the preceding stage, we estimate the global shrinkage parameter by maximum marginal likelihood,
\begin{equation}
\widehat\theta
\in
\arg\max_{\theta\in\Theta}\,
\pi(y\mid\theta),
\qquad
\pi(y\mid\theta)
=
\int \pi(y\mid\beta)\pi(\beta\mid\theta)\,d\beta,
\label{eq:eb-target}
\end{equation}
over a compact interval \(\Theta\subset(0,\infty)\).\ Once \(\widehat\theta\) has been obtained, it is held fixed for the subsequent MAP and posterior calculations.\ The computational difficulty in \eqref{eq:eb-target} is not the one-dimensional optimisation itself, but evaluating the marginal-likelihood score without access to the integral in closed form.

SAPG \citet{vidal2020part1,deBortoli2020part2} addresses this through a Fisher-identity representation of the score and a stochastic-approximation update whose expectation is estimated by a short proximal Markov chain.\ To make the quantity used in this paper explicit, let \(C\) denote the support of the prior, with intrinsic dimension \(d\), and let \(\mu_d\) denote the induced \(d\)-dimensional Lebesgue measure on \(C\).\ We write
\[
Z_C(\theta)
=
\int_C \exp\{-\theta g(\beta)\}\,d\mu_d(\beta)
\]
for its normalising constant.\ The marginal-likelihood derivative satisfies
\begin{equation}
\frac{\partial}{\partial\theta}\log \pi(y\mid\theta)
=
-\mathbb E_\theta\!\left[g(\beta)\mid y\right]
-\frac{\partial}{\partial\theta}\log Z_C(\theta).
\label{eq:eb-fisher}
\end{equation}
For a one-homogeneous penalty, \(g(t\beta)=t g(\beta)\), on a \(d\)-dimensional linear support, the change of variables \(u=\theta\beta\) gives
\[
Z_C(\theta)=\theta^{-d}Z_C(1).
\]
Hence, in logarithmic coordinates \(\eta=\log\theta\),
\begin{equation}
\frac{\partial}{\partial\eta}
\log \pi(y\mid e^\eta)
=
d-\theta\,
\mathbb E_\theta\!\left[g(\beta)\mid y\right].
\label{eq:sapg-score}
\end{equation}
For the unconstrained Laplace prior, \(C=\mathbb R^p\), \(d=p\), \(g(\beta)=\|\beta\|_1\), and \(Z_C(\theta)=(2/\theta)^p\).

At iteration \(k\), a short MYULA chain at the current \(\theta_k=e^{\eta_k}\) provides states \(\{\beta_k^{(\ell)}\}_{\ell=1}^{m_k}\).\ Writing
\[
\overline g_k
=
\frac{1}{m_k}\sum_{\ell=1}^{m_k}g(\beta_k^{(\ell)}),
\qquad
\widehat\Delta_k
=
d-\theta_k\overline g_k,
\]
we use the projected Robbins--Monro recursion
\begin{equation}
\eta_{k+1}
=
\Pi_{\log\Theta}
\left\{
\eta_k+\rho_k\widehat\Delta_k
\right\},
\qquad
\rho_k=\frac{c}{k+k_0},
\label{eq:sapg-update}
\end{equation}
followed by a tail/Polyak--Ruppert average of the resulting parameter path.\ Algorithm~\ref{alg:sapg} summarises the calculation.\

\begin{algorithm}[H]
\caption{SAPG calibration of a homogeneous Laplace scale}
\label{alg:sapg}
\begin{algorithmic}[1]
\Require \(y,X,\widehat\sigma^2\), geometry-specific \(f\), intrinsic dimension \(d\), initial \(\theta_0\), bounds \(\Theta\), steps \(\rho_k\).
\State Initialise a chain state \(\beta_0\) and \(\eta_0=\log\theta_0\).
\For{\(k=0,\ldots,K-1\)}
  \State At \(\theta_k=e^{\eta_k}\), advance a short MYULA chain and collect \(m_k\) states.
  \State \(\overline g_k\gets m_k^{-1}\sum_{\ell=1}^{m_k}g(\beta_k^{(\ell)})\).
  \State \(\widehat\Delta_k\gets d-\theta_k\overline g_k\).
  \State \(\eta_{k+1}\gets\Pi_{\log\Theta}\{\eta_k+\rho_k\widehat\Delta_k\}\).
\EndFor
\State Return a tail/Polyak--Ruppert average as \(\widehat\theta\).
\end{algorithmic}
\end{algorithm}
\FloatBarrier

The simple score in \eqref{eq:sapg-score} is a consequence of one-homogeneity, not a requirement of the broader optimisation--sampling framework.\ For example, an elastic net prior with fixed mixing parameter still has an explicit proximal map, so the FISTA and proximal MCMC steps remain straightforward.\ Its empirical Bayes update can use the general Fisher/SAPG identity \eqref{eq:eb-fisher} with the appropriate derivative of the prior normalising constant in place of the term \(d/\theta\);\ if both elastic net penalty parameters are estimated, the stochastic-approximation update becomes multivariate, but can still be treated by other variants of SAPG already outlined in \citet{vidal2020part1}.

\paragraph{Hard homogeneous constraints}
For a hard linear restriction
\[
C_0=\{\beta:A\beta=0\},
\qquad
d=p-\operatorname{rank}(A),
\]
the support is a linear subspace, so the same scaling argument applies with its \emph{intrinsic} dimension \(d\).\ In particular, under the sum-zero condition \(\bm 1^\top\beta=0\), the correct score in \eqref{eq:sapg-score} uses \(d=p-1\), not \(p\).\ The same intrinsic dimension enters the natural scale-matching initialisation
\begin{equation}
\theta_0=\frac{d}{g(\beta_{\rm ref})}.
\label{eq:theta0}
\end{equation}
Here \(\beta_{\rm ref}\) is a preliminary reference fit used only to set the initial shrinkage scale;\ the experiment-specific choices are given in the supplement.\ The short MYULA chain used inside SAPG is run within \(C_0\), using the projected smooth gradient, projected Gaussian innovation and constrained proximal map;\ the explicit update and implementation details are given in the supplementary material.\ A paired numerical verification reported there, deliberately replaces \(d=p-1\) by the ambient value \(p\) in the sum-zero example.\

\paragraph{Nonhomogeneous affine information}
The same argument cannot simply be transferred to an affine slice
\[
C_b=\{\beta:A\beta=b\},
\qquad b\neq0.
\]
Indeed, under the rescaling \(u=\theta\beta\), the constraint becomes \(Au=\theta b\), so the integration domain itself depends on \(\theta\) and \(Z_{C_b}(\theta)\) does not factor as \(\theta^{-d}Z_{C_b}(1)\).\ A dimension-only contribution to the SAPG score is therefore no longer available.\ Rather than define the Laplace prior directly on this moving affine slice, we retain the prior on ambient \(\mathbb R^p\) and impose the nonhomogeneous relation softly through the Gaussian pseudo-observation
\begin{equation}
f_{\tau_c}(\beta)
=
\frac{1}{2\sigma^2}\|y-X\beta\|_2^2
+
\frac{1}{2\tau_c^2}\|A\beta-b\|_2^2.
\label{eq:softaffine}
\end{equation}
The prior normalising constant then remains proportional to \(\theta^{-p}\), so \eqref{eq:sapg-score} continues to use \(d=p\), while \(\tau_c\) controls both the strength of the affine information and the additional curvature introduced by the quadratic affine term.\ Section~\ref{study: two} studies these two consequences separately.

For the unconstrained and soft affine cases we likewise use \(d=p\) in \eqref{eq:theta0};\ the choices of \(\beta_{\rm ref}\), projection bounds and averaging windows used in the experiments are recorded in the supplementary material.\ We note that the exact score \eqref{eq:sapg-score} belongs to the nonsmoothed model, whereas its posterior expectation is estimated with the inexact MYULA kernel described in Section~\ref{subsec:MY}.\ Consequently, \(\widehat\theta\) can inherit some dependence on the Moreau parameter \(\lambda\).\ We retain the nonsmoothed marginal likelihood as the empirical Bayes target and assess this numerical dependence separately.

\subsection{MAP estimation and role in selection}

The nonsmoothed MAP has two roles.\ It is the primary point estimator under the fitted empirical Bayes model, and its joint sparse configuration provides the candidate set and magnitude anchor for the later selection rule.\ The posterior gates therefore add uncertainty information without replacing the MAP as the coefficient estimator.\ Our index-tracking application \citet{roxanas2025index} is an example where the MAP is more explicitly an intermediate decision vector because portfolio weights or trades must still be converted into an implementable constrained action;\ here the MAP itself remains the reported point estimate, but the applicability of the method is broader.

Conditional on $(\widehat\sigma^2,\widehat\theta)$, the primary estimator is
\begin{equation}
\widehat\beta_{\rm MAP}
\in
\arg\min_\beta
\{f(\beta)+\widehat\theta\|\beta\|_1\},
\label{eq:map}
\end{equation}
where \(f\) is the active smooth term defined in subsection~\ref{subsec:MY}.\ We compute \eqref{eq:map} by FISTA.\ If $\nabla f$ is $L_f$-Lipschitz, the unconstrained and soft affine iterations use $\gamma=1/L_f$ and, at the extrapolated point $z^k$,
\begin{equation}
  \beta^{k+1}
  =\prox_{\gamma\widehat\theta g}
  \left(z^k-\gamma\nabla f(z^k)\right).
  \label{eq:fista-prox}
\end{equation}
For $g(\beta)=\|\beta\|_1$, this is coordinatewise soft thresholding at level $\gamma\widehat\theta$.

The empirical Bayes estimate $\widehat\theta$ is obtained with a MYULA approximation and can therefore depend on the Moreau parameter $\lambda$.\ Once $\widehat\theta$ is fixed, however, FISTA in \eqref{eq:map} optimises the original nonsmoothed objective.\ For comparison, the smoothed posterior has mode
\begin{equation}
  \widehat\beta_{\mathrm{MAP},\lambda}
  \in\argmin_\beta
  \left\{f(\beta)+h_{\widehat\theta}^\lambda(\beta)\right\}.
  \label{eq:smoothed-map}
\end{equation}
In general $\widehat\beta_{\mathrm{MAP},\lambda}\neq\widehat\beta_{\mathrm{MAP}}$.\ We report the nonsmoothed MAP and use the smoothed mode only to audit the approximation.\ In particular, we monitor
\begin{equation}
  D_2(\lambda)
  =\frac{\|\widehat\beta_{\mathrm{MAP},\lambda}-\widehat\beta_{\mathrm{MAP}}\|_2}
  {1+\|\widehat\beta_{\mathrm{MAP}}\|_2},
  \qquad
  D_\infty(\lambda)
  =\|\widehat\beta_{\mathrm{MAP},\lambda}-\widehat\beta_{\mathrm{MAP}}\|_\infty.
  \label{eq:map-discrepancy}
\end{equation}

For the hard sum-zero constraint \(C_0=\{\beta:\bm1^\top\beta=0\}\), the constrained proximal map is
\begin{equation}
\prox^{C_0}_{t\theta\|\cdot\|_1}(v)
=\argmin_{u\in C_0}
\left\{\frac12\|u-v\|_2^2+t\theta\|u\|_1\right\}.
\end{equation}
Its KKT representation is
\begin{equation}
  u_j=S_{t\theta}(v_j-\nu),
  \qquad
  \sum_{j=1}^p S_{t\theta}(v_j-\nu)=0,
  \label{eq:sumzero-prox}
\end{equation}
where \(S_a(x)=\operatorname{sign}(x)(|x|-a)_+\) is the scalar soft-thresholding operator and the scalar $\nu$ is obtained by bisection \citet[e.g., Section~5.1]{zhou2024}.\ FISTA then operates directly in \(C_0\):
\begin{equation}
  \beta^{k+1}
  =\prox^{C_0}_{\gamma\widehat\theta g}
  \left(z^k-\gamma P_{C_0}\nabla f_0(z^k)\right),
  \label{eq:constrained-fista}
\end{equation}
with
\begin{equation}
  P_{C_0}=I-\frac{1}{p}\bm1\bm1^\top.
\end{equation}

\subsection{Posterior-informed sparse decisions}\label{subsec: gates}

A continuous Laplace posterior does not contain inclusion indicators, and MAP nonzero status alone need not imply practical relevance.\ We therefore separate shrinkage from the final support decision, following the general idea of fitting a continuous model first and sparsifying a posterior summary afterward \citet{bondell2012,hahn2015,piironen2020}.\ The nonsmoothed MAP supplies the joint magnitude anchor, while MYULA provides posterior-scale information from the smoothed approximation.

From retained MYULA draws, we define
\begin{equation}
  \widehat s_j(\lambda)
  =\operatorname{sd}_{\pi_{\widehat\theta,\lambda}}
  (\beta_j\mid y),
  \qquad j=1,\ldots,p.
  \label{eq:post-sd}
\end{equation}
We use the robust global scale
\begin{equation}
  \tau_{\mathrm{post}}(\lambda;k)
  =k\,\operatorname{median}_{1\le j\le p}\widehat s_j(\lambda),
  \label{eq:taupost}
\end{equation}
which provides a common posterior scale in coefficient units while preventing a small number of coordinates with very large marginal posterior standard deviations from setting the threshold for all variables.\ We define the activation probability
\begin{equation}
  \widehat\pi_j(\lambda;k)
  =\Pr_{\pi_{\widehat\theta,\lambda}}
  \left(
  |\beta_j|\ge\tau_{\mathrm{post}}(\lambda;k)
  \mid y
  \right),
  \label{eq:activation}
\end{equation}
estimated by Monte Carlo frequencies.\

A marginal activation probability can be large for two different reasons:\ the posterior may be centred away from zero, or it may simply be very spread out.\ To make the latter point explicit, suppose for illustration that the marginal posterior of one coefficient is approximately
\[
\beta_j\mid y \approx \mathcal N(\mu_j,s_j^2).
\]
For a fixed practical threshold $\tau>0$,
\[
\Pr(|\beta_j|\ge\tau\mid y)
=
\Phi\!\left(\frac{-\tau-\mu_j}{s_j}\right)
+
1-\Phi\!\left(\frac{\tau-\mu_j}{s_j}\right),
\]
where $\Phi$ is the standard normal distribution function.\ If $\mu_j=0$, this becomes
\[
2\left\{1-\Phi\!\left(\frac{\tau}{s_j}\right)\right\},
\]
which increases with $s_j$ (and hence with the posterior variance $s_j^2$) for fixed $\tau$.\ These are ``diffuse'' in the sense that a weakly identified coordinate can place substantial posterior mass beyond $\pm\tau$ because its marginal variance is large, even when it is centred near zero.\ The median in \eqref{eq:taupost} limits the influence of a small number of such coordinates on the common threshold, while the MAP magnitude gate below prevents marginal spread alone from triggering selection.

The selected support is
\begin{equation}
\widehat S_\lambda(k,\pi_\star)
  =\left\{
  j:
  |\widehat\beta_{\mathrm{MAP},j}|\ge\tau_{\mathrm{post}}(\lambda;k),
  \quad
  \widehat\pi_j(\lambda;k)\ge\pi_\star
  \right\}.
\label{eq:gate}
\end{equation}
We report
\[
k\in\{2,2.5,3\},
\qquad
\pi_\star\in\{0.5,0.75,0.9\},
\]
with $(2.5,0.75)$ as the reference setting.\ The MAP gate is tied to the nonsmoothed fitted model, whereas the posterior scale and activation probability depend on the Moreau approximation.

Conditional on passing the MAP magnitude gate, the probability threshold has a simple coordinatewise decision-theoretic interpretation.\ Let
\begin{equation}
  E_j(\lambda;k)
  =\{|\beta_j|\ge\tau_{\mathrm{post}}(\lambda;k)\}
\end{equation}
denote the event that coefficient $j$ is practically active, and let $a_j\in\{0,1\}$ denote the decision to exclude ($a_j=0$) or select ($a_j=1$) that coordinate.\ Let $c_{\mathrm{FP}}>0$ be the loss assigned to selecting a practically inactive coefficient and $c_{\mathrm{FN}}>0$ the loss assigned to excluding a practically active one.\ With $\ind(\cdot)$ denoting the indicator function, consider
\begin{equation}
  L(a_j,E_j)
  =c_{\mathrm{FP}}a_j\ind(E_j^c)
  +c_{\mathrm{FN}}(1-a_j)\ind(E_j).
  \label{eq:decision-loss}
\end{equation}
The posterior expected losses of selection and exclusion are therefore
\[
c_{\mathrm{FP}}\Pr(E_j^c\mid y)
\qquad\text{and}\qquad
c_{\mathrm{FN}}\Pr(E_j\mid y),
\]
respectively.\ The Bayes action selects $j$ when
\begin{equation}
  \Pr(E_j\mid y)
  \ge
  \frac{c_{\mathrm{FP}}}{c_{\mathrm{FP}}+c_{\mathrm{FN}}}.
  \label{eq:bayes-threshold}
\end{equation}
Thus $\pi_\star$ can be read as a relative false-inclusion/false-exclusion cost threshold.\ For example, $\pi_\star=0.75$ corresponds to assigning a false inclusion three times the loss of a false exclusion at this coordinatewise decision stage.

Because the terminal decision depends on Monte Carlo estimates of \eqref{eq:taupost} and \eqref{eq:activation}, we supplement standard autocorrelation and effective-sample-size diagnostics with a decision-level quantity.\ For retained draw $m$, define the binary activation sequence
\[
I_j^{(m)}(\lambda;k)
=
\ind\!\left\{
|\beta_j^{(m)}|\ge\tau_{\mathrm{post}}(\lambda;k)
\right\}.
\]
Then $\widehat\pi_j(\lambda;k)$ is the sample mean of this sequence, and $\operatorname{MCSE}\{\widehat\pi_j(\lambda;k)\}$ is the Monte Carlo standard error of that mean, accounting for serial dependence through the effective sample size.\ We define
\begin{equation}
D_j(\lambda;k,\pi_\star)
=
\frac{|\widehat\pi_j(\lambda;k)-\pi_\star|}
{\operatorname{MCSE}\{\widehat\pi_j(\lambda;k)\}}.
\label{eq:D}
\end{equation}
We report the minimum $D_j$ over all coordinates and, more importantly, over coordinates that pass the MAP magnitude gate.\ If $I_j^{(m)}(\lambda;k)$ is constant over the retained draws, its estimated variance and hence its estimated MCSE are zero.\ For the interior probability thresholds used here, the empirical activation probability is then either zero or one, so the retained run exhibits no Monte Carlo ambiguity about which side of the probability gate that coordinate lies on.\ This should not be interpreted as posterior certainty; it is a statement about Monte Carlo resolution conditional on the sampled run and is considered together with the usual mixing diagnostics.

\subsection{Geometry-aware preconditioning}
\label{sec:preconditioning}

The scalar-step MYULA update introduced above is most effective when the posterior curvature is reasonably balanced across directions.\ This need not hold in regression problems:\ the likelihood may be poorly scaled across coordinates, while a strong soft affine relation can introduce a small number of directions whose curvature is much larger than that of the remaining parameter space.\ Both effects can restrict the admissible timestep and degrade mixing in directions that are not themselves stiff.\ We therefore use preconditioning as a numerical device matched to the source of the anisotropy.

Let \(U_{\theta,\lambda}\) denote the smoothed negative log-posterior and let \(M\) be a fixed symmetric positive-definite matrix.\ The corresponding preconditioned Langevin diffusion is
\[
    d\beta_t
    =
    -M\nabla U_{\theta,\lambda}(\beta_t)\,dt
    +\sqrt{2M}\,dW_t,
\]
which has the same invariant distribution as the unpreconditioned diffusion.\ Its Euler discretisation is
\[
    \beta^{(k+1)}
    =
    \beta^{(k)}
    -\delta M\nabla U_{\theta,\lambda}\!\left(\beta^{(k)}\right)
    +\sqrt{2\delta M}\,\xi^{(k+1)},
    \qquad
    \xi^{(k+1)}\sim\mathcal N(0,I_p).
\]
Thus a constant preconditioner changes the geometry in which the target is explored, without changing the intended smoothed posterior.\ The timestep must nevertheless be retuned in the transformed geometry.\ For a quadratic smooth term with Hessian \(H_f\), we denote the corresponding smooth-gradient Lipschitz constant in the preconditioned geometry by
\[
L_{f,M}
=
\lambda_{\max}\!\left(M^{1/2}H_fM^{1/2}\right).
\]

\paragraph{Diagonal scaling.}
When poor conditioning is distributed across coordinates, one can use a diagonal, or Jacobi, approximation to the Hessian of the smooth term.\ Writing \(H_f=\nabla^2 f\) for the Hessian of the quadratic smooth term, we take
\[
    D_\varepsilon
    =
    \operatorname{diag}(H_f)+\varepsilon I_p ,\qquad \varepsilon>0,
    \qquad
    M_{\rm J}=D_\varepsilon^{-1},
\]
where \(\varepsilon\) is a small ridge term that prevents zero or very small diagonal entries from producing unstable inverse rescaling.\ Up to an overall scalar normalisation that can be absorbed into \(\delta\), this rescales coordinates according to the diagonal entries of \(H_f\) and is inexpensive to construct even when \(p\) is large.\ It is useful when the main numerical difficulty is heterogeneous marginal scaling; being diagonal, it is not intended to remove strong off-diagonal dependence completely.

\paragraph{Low-rank directional correction.}
A different geometry arises from a soft affine relation.\ For a single condition \(a^\top\beta=b\), the pseudo-observation adds
\[
    \frac{1}{2\tau_c^2}(a^\top\beta-b)^2
\]
to the smooth potential and hence contributes the rank-one Hessian term \(\tau_c^{-2}aa^\top\).\ Let
\[
    u=\frac{a}{\|a\|_2},
    \qquad
    P_{\parallel}=uu^\top,
    \qquad
    P_{\perp}=I-P_{\parallel}.
\]
When the largest eigenvalue of the smooth Hessian is driven primarily by the soft affine term, it is useful to distinguish the overall smooth-gradient Lipschitz constant from the corresponding constant within the tangent space.\ For this quadratic problem, define
\[
    L_f=\lambda_{\max}(H_f),
    \qquad
    L_{\perp}
    =
    \sup_{\substack{v\perp u\\ \|v\|_2=1}}
    v^\top H_f v.
\]
Thus \(L_f\) is the largest eigenvalue of the smooth Hessian, while \(L_{\perp}\) is its largest Rayleigh quotient over directions tangent to the affine hyperplane.\ We then use
\[
    M_\phi
    =
    P_{\perp}+\phi P_{\parallel},
    \qquad
    \phi
    =
    \min\left\{1,\frac{L_{\perp}}{L_f}\right\}.
\]
The update therefore leaves tangent directions unchanged while damping movement in the single stiff direction enough to bring its numerical scale closer to the rest of the problem.\ The same idea extends to several affine relations by replacing \(P_{\parallel}\) with the projector onto the relevant low-dimensional constraint subspace.

There is one additional interaction with Moreau smoothing.\ If the soft affine Hessian term materially increases \(L_f\), the default choice \(\lambda=1/L_f\) also makes the Moreau approximation unnecessarily sharp in the better-scaled tangent directions.\ In that case we decouple the two scales and use the tangent-space constant \(L_{\perp}\) to set the Moreau parameter, while the timestep is chosen from the Lipschitz bound in the preconditioned geometry.\ 

Hence the two preconditioners address different mechanisms:\ diagonal scaling treats broadly distributed coordinatewise imbalance, whereas the projector construction treats identifiable low-rank stiffness.\ We assess the effect of preconditioning through mixing, effective sample size and the Monte Carlo precision of the posterior quantities entering the final decision rule.


\section{Numerical studies}
\label{sec:studies}

Study~1 tests the full unconstrained workflow and, in particular, whether calibration and Monte Carlo choices alter the final sparse decision in the $n>p$ and $p>n$ regimes.\ Study~2 isolates the statistical and numerical consequences of soft affine information.\ Study~3 asks whether the resulting posterior uncertainty and sparse decision remain credible on a real regression problem with correlated predictors.\ The supplement contains the hard-constraint calibration check and the observation-scale/Moreau sensitivity study.\ Unless stated otherwise, selection results use $(k,\pi_\star)=(2.5,0.75)$ as the reference setting and report the full grid as a sensitivity analysis.

\subsection{Study 1: End-to-end sparse regression}
\label{study: one}

Study~1 evaluates the complete workflow under a known sparse ground truth, with emphasis on how upstream calibration and posterior Monte Carlo precision propagate to the final support decision.\ We use the same signal pattern in an overdetermined and an underdetermined regime and report likelihood-scale calibration, MAP estimation and prediction, sparse selection, and decision-level Monte Carlo precision.

Rows of the Gaussian design are generated with AR(1) covariance \(\Sigma_{jk}=0.4^{|j-k|}\), after which each realised predictor column is rescaled to \(\|X_j\|_2=\sqrt n\).\ The response follows \(y=X\beta^\star+\varepsilon\), with \(\varepsilon\sim\mathcal N(0,I_n)\), so that \(\sigma=1\).\ There are ten nonzero coefficients, with alternating signs and decreasing magnitudes, rescaled so that the population signal-to-noise ratio satisfies
\[
\frac{{\beta^\star}^{\!\top}\Sigma\beta^\star}{\sigma^2}=3.
\]
We consider
\[
(n,p)=(500,200)
\qquad\text{and}\qquad
(n,p)=(200,500),
\]
with \(R=100\) independently generated datasets in each regime; the additional predictors in the high-dimensional regime are null.\ The decision rule is evaluated over \(k\in\{2,2.5,3\}\) and \(\pi_\star\in\{0.50,0.75,0.90\}\), with \((2.5,0.75)\) used only as a reference operating point.\ SAPG uses \(4000\) iterations with tail averaging from iteration \(2000\).\ The initial calculation retained \(4000\) MYULA draws after burn-in and was subsequently refined to \(40{,}000\) retained draws with all upstream quantities fixed.\ The full settings and the complete decision grid are given in the supplementary material.\ Throughout the numerical studies, TPR denotes true-positive rate, FDR false-discovery rate, and support RMSE the root mean squared coefficient error restricted to the true support.

The two regimes use the two scale-estimation branches of subsection~\ref{subsec:noise}.\ When \(n>p\), the classical residual estimator is centred close to the truth:\ across the \(R=100\) datasets,
\[
\operatorname{mean}(\widehat\sigma/\sigma)=0.9979,
\qquad
\operatorname{sd}(\widehat\sigma/\sigma)=0.0415.
\]

The high-dimensional branch required more care.\ In an initial five-dataset pilot, the raw scaled Lasso estimate was materially inflated, with mean \(\widehat\sigma_{\rm SL}=1.514\), and the initially narrow SAPG projection interval also produced boundary hits.\ Widening the projection interval removed this numerical artefact, but selection remained weak when the inflated scale was retained.\ Replacing the estimated scale by the true \(\sigma=1\) recovered much of the lost sensitivity and improved prediction performance.\ This separated the two effects and motivated a feasible correction to the scale estimate:\ we use the scaled Lasso only to identify a provisional support, refit OLS on that support, and estimate \(\sigma^2\) from the refitted residuals as in \eqref{eq:postols-sigma}.\ In the same five datasets, this reduced the mean estimated scale from \(1.514\) to \(1.138\), with behaviour much closer to the true-noise comparison.

Accordingly, the reported \(R=100\) experiment uses the post-selection OLS scale, while the scaled Lasso coefficients themselves play no role beyond defining the provisional support.\ In the \(p>n\) datasets, the raw scaled Lasso scale has mean \(1.4820\), whereas the post-OLS estimate has mean \(1.1173\) and standard deviation \(0.0776\).\ The remaining upward displacement is consistent with provisional supports that omit weak signals; the supplementary analysis records their composition explicitly.\ This sequence is important for the interpretation of the later studies as it clarifies the role of the observation-scale estimate in the full workflow.\

Table~\ref{tab:study1main} summarises the output at the reference decision setting, using the final \(40{,}000\)-draw posterior calculations for all selection quantities.\ Relative \(L_2\) error and true-support RMSE refer to the nonsmoothed MAP, while prediction risk is evaluated under the population design covariance.

\begin{table}[H]
\centering
\caption{Study~1 results at the reference setting \((k,\pi_\star)=(2.5,0.75)\).\ Parentheses give standard errors over \(R=100\) datasets;\ selection uses the \(40{,}000\)-draw MYULA run.}
\label{tab:study1main}
\small
\resizebox{\textwidth}{!}{%
\begin{tabular}{cccccccccc}
\toprule
Regime & \(\widehat\sigma/\sigma\) & \(\widehat\theta\) &
Rel.\ \(L_2\) & Supp.\ RMSE & Pred.\ risk & TPR & FDR & Size & Exact\\
\midrule
\(n=500,p=200\) &
0.998 (.004) & 15.711 (.041) & 0.215 (.003) & 0.065 (.001) & 0.130 (.003) &
0.989 (.003) & 0.006 (.002) & 9.96 (.04) & 0.830 (.038)\\
\(n=200,p=500\) &
1.117 (.008) & 17.334 (.067) & 0.343 (.004) & 0.165 (.003) & 0.347 (.009) &
0.429 (.012) & 0.000 (.000) & 4.29 (.12) & 0.000 (.000)\\
\bottomrule
\end{tabular}
}%
\end{table}

The low-dimensional regime, \(n>p\), behaves as a well-informed sparse regression problem.\ The noise scale is accurately calibrated, estimation and prediction errors remain moderate, and the reference rule recovers on average \(98.9\%\) of the true variables with FDR \(0.0065\); the exact ten-variable support is recovered in \(83\%\) of datasets.\ The high-dimensional regime, \(p\ge n\), is qualitatively different.\ Estimation and prediction remain meaningful, but the final decision at the reference setting, $(k,\pi_\star)=(2.5,0.75)$, is conservative:\ it selects \(4.29\) variables on average with TPR \(0.429\), but no false positives are observed in the \(100\) replications.\ Signal-wise selection frequencies decrease almost monotonically with true coefficient magnitude, from \(0.96\) for the largest signal to zero for the smallest.\ The loss of exact support recovery is therefore driven by omissions of progressively weaker signals, rather than by indiscriminate inclusion of zero coordinates.

\begin{figure}[htbp]
\centering
\includegraphics[width=0.68\textwidth]{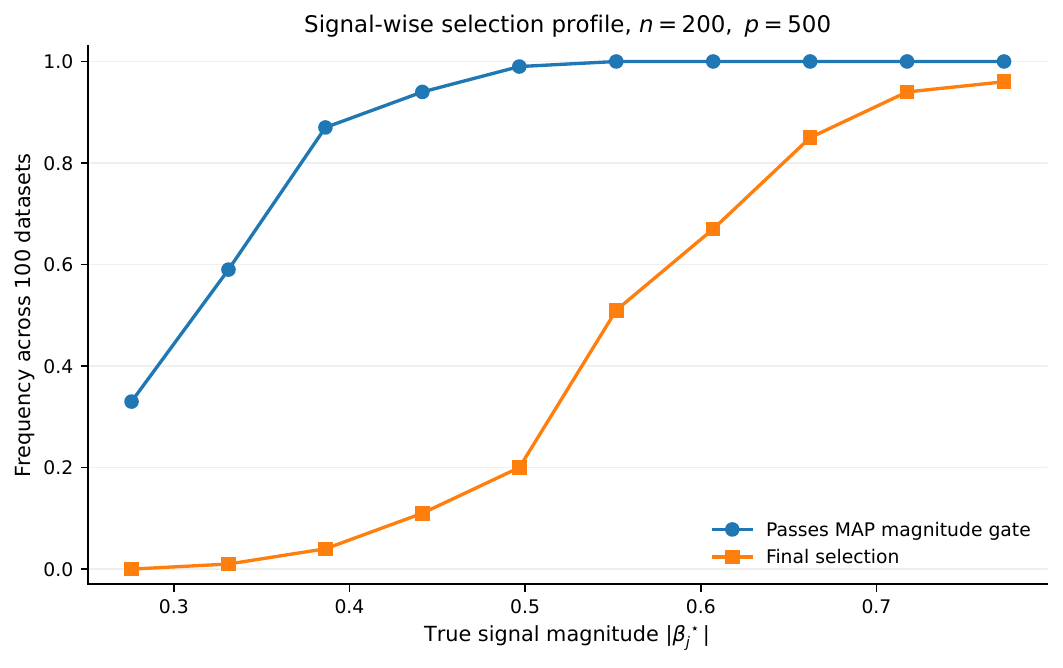}
\caption{Signal-wise behaviour at the reference decision rule in the high-dimensional regime. For each true signal, the figure shows the frequency with which it passes the MAP magnitude gate and the frequency with which it is retained by the complete MAP-plus-posterior rule.}
\label{supp:fig:study1_signalwise}
\end{figure}

Figure~\ref{supp:fig:study1_signalwise} makes the two-stage mechanism explicit in the high-dimensional regime:\ MAP eligibility remains high for moderate signals, while the posterior activation gate progressively removes weaker effects.

The full \(3\times3\) decision grid is obtained by post-processing the same posterior sample; neither the model nor the empirical Bayes calibration is refitted when \(k\) or \(\pi_\star\) changes.\ In the low-dimensional regime, the grid displays the expected sensitivity--specificity tradeoff:\ the permissive setting \((2,0.50)\) has TPR \(1\) and FDR \(0.105\), while more stringent settings almost eliminate false discoveries with only a modest loss of sensitivity.\ In the high-dimensional regime, FDR is zero over almost the whole grid;\ TPR ranges from \(0.675\) at \((2,0.50)\) to \(0.229\) at \((3,0.90)\).\ Thus, the two decision parameters mainly control how much weak-signal sensitivity is retained once the problem becomes information-limited.

Increasing the retained MYULA sample from \(4000\) to \(40{,}000\) draws separates the statistical behaviour above from finite Monte Carlo precision.\ Median coordinate ESS increases from \(212\) to \(2093\) in the \(n>p\) regime and from \(134\) to \(1269\) in the \(p>n\) regime, while aggregate TPR and FDR move by less than one percentage point.\ Let \(\widehat S_{4k}\) and \(\widehat S_{40k}\) denote the selected sets obtained from the shorter and longer posterior runs.\ These sets are identical in \(96\%\) of low-dimensional replicates and \(74\%\) of high-dimensional replicates; the mean values of \(|\widehat S_{4k}\triangle\widehat S_{40k}|\), the number of coordinates selected in exactly one of the two runs, are only \(0.04\) and \(0.33\), respectively.

The decision-aware diagnostic \(D_j\), defined in \eqref{eq:D}, measures the distance between the estimated activation probability for coefficient \(j\) and the probability threshold \(\pi_\star\), in Monte Carlo standard-error
units. For each replicate, we restrict attention to coordinates that pass the
MAP magnitude gate and record
\[
D_{\min,\mathrm{elig}}
=
\min_{j:\,|\widehat\beta_{\mathrm{MAP},j}|
\geq \tau_{\mathrm{post}}}
D_j.
\]
Thus, \(D_{\min,\mathrm{elig}}\) measures how close the most
decision-sensitive MAP-eligible coefficient is to the posterior-probability boundary:\ a small value means that at least one eligible coefficient has an activation probability close to \(\pi_\star\) relative to its Monte Carlo uncertainty.\ The median of this quantity increases from \(7.03\) to \(27.87\) in the \(n>p\) regime and from \(0.81\) to \(2.16\) in the \(p>n\) regime.\ 
%

\begin{figure}[H]
\centering
\includegraphics[width=0.68\textwidth]{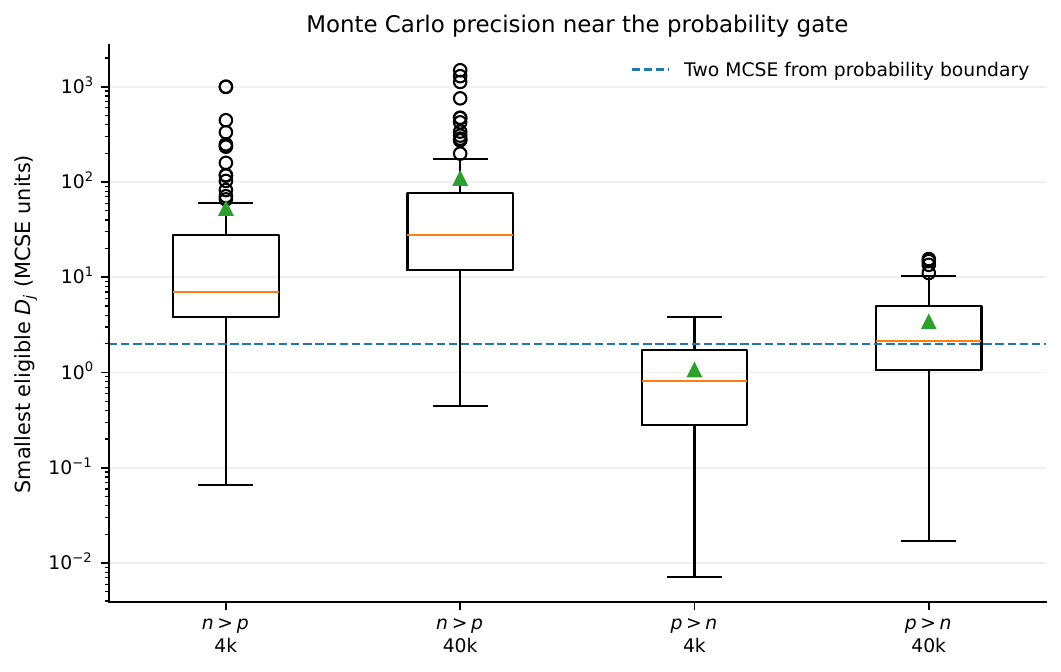}
\caption{Monte Carlo precision of the posterior-probability gate under the \(4000\)- and \(40{,}000\)-draw calculations. The plotted quantity is the smallest \(D_j\) among MAP-eligible coordinates in each replicate; the dashed line marks two Monte Carlo standard errors from the probability threshold.}
\label{supp:fig:study1_decisionprecision}
\end{figure}

Even after \(40{,}000\) draws, \(46\%\) of the high-dimensional replicates contain at least one MAP-eligible coordinate within two MCSE units of the activation threshold.\ Thus, the aggregate recovery rates are stable, while individual high-dimensional inclusion decisions can remain close enough to the probability gate for Monte Carlo precision to matter.\ Increasing the retained sample tenfold was computationally inexpensive with MYULA in these experiments; environment-specific timings are reported in the supplement.\
Figure~\ref{supp:fig:study1_decisionprecision} visualises the same distinction:\ the longer posterior run moves the most decision-sensitive eligible coordinates farther from the probability boundary in Monte Carlo standard-error units.

In conclusion, Study~1 establishes the baseline behaviour used by the remaining experiments.\ It shows that the complete pipeline is well behaved in the \(n>p\) regime, that the high-dimensional difficulty is primarily conservative loss of weak signals, and that nuisance-scale calibration and posterior Monte Carlo error can be diagnosed separately.\

\FloatBarrier

\subsection{Study 2: Soft affine information and numerical geometry} \label{study: two}

Study~2 isolates the second part of the methodology:\ how affine information affects both empirical Bayes calibration and posterior computation.\ In our index-tracking application \citet{roxanas2025index}, construction uses a soft sum-to-one condition and rebalancing uses hard self-financing changes with $\mathbf 1^\top\Delta w=0$.\ Here we isolate the soft nonhomogeneous case in a controlled regression experiment.\ We examine how stronger sum-to-one information changes calibration, estimation and selection, and how the same term changes the geometry seen by the Langevin sampler.\ The hard homogeneous calibration is checked separately in the supplement.

We use \(R=100\) independently generated datasets with \(n=200\), \(p=100\), known \(\sigma=1\), and eight nonzero coefficients satisfying \(\mathbf 1^\top\beta^\star=1\).\ The smooth potential is augmented by the Gaussian pseudo-observation term
\[
\frac{1}{2\tau_c^2}\bigl(\mathbf 1^\top\beta-1\bigr)^2 
\]
that encodes the constraint.\ Writing \(L_0=\lambda_{\max}(X^\top X)\), we index the strength of the affine information by
\[
\kappa_c
=
\frac{p}{\tau_c^2 L_0},
\qquad
\kappa_c\in\{0,0.1,1,10\}.
\]
This parameter has a direct geometric interpretation:\ the rank-one affine Hessian
\(\tau_c^{-2}\mathbf 1\mathbf 1^\top\) has nonzero eigenvalue \(p/\tau_c^2=\kappa_cL_0\).\ Thus \(\kappa_c\) compares the curvature introduced in the sum direction with the baseline likelihood-curvature scale; \(\kappa_c=0\) corresponds to omitting the pseudo-observation.\ Because the relation is imposed softly rather than by restricting the prior support, the prior remains on the ambient space and the SAPG score continues to use \(d=p\) throughout.\ The full simulation settings and decision grids are reported in the supplement.

The empirical Bayes scale changes only moderately as the affine information strengthens, from mean \(\widehat\theta=9.772\) at \(\kappa_c=0\) to \(10.346\) at \(\kappa_c=10\).\ The effect on adherence to the supplied relation is much larger:\ the mean posterior value of \(|\mathbf 1^\top\beta-1|\) decreases from \(0.685\) to \(0.112\).\ Point estimation and selection improve more modestly.\ MAP relative \(L_2\) error falls from \(0.2557\) to \(0.2482\), while at the reference decision rule, 
\((k,\pi_\star)=(2.5, 0.75)\), mean TPR increases from \(0.8025\) to \(0.8200\);\ FDR remains below \(0.003\) at both endpoints.\ Table~\ref{tab:study2} summarises the final posterior calculation used for each value of \(\kappa_c\).\ Here Med.\ ESS is the ensemble median of the coordinatewise median effective sample size, Weak-tangent ESS is the ensemble median of the effective sample size after projection onto the least-curved tangent-space eigendirection, and \(D_{\rm elig}\) is the median, across datasets, of the smallest \(D_j\) among MAP-eligible coordinates.

\begin{table}[htbp]
\centering
\caption{Study~2 results at \((k,\pi_\star)=(2.5,0.75)\).\ The \(\kappa_c=10\) row uses the geometry-aware calculation;\ the baseline scalar-step calculation is reported in the supplement.}
\label{tab:study2}
\small
\resizebox{\textwidth}{!}{%
\begin{tabular}{lcccccccc}
\toprule
\(\kappa_c\) & \(\widehat\theta\) & MAP rel.\ \(L_2\) &
\(E|\mathbf 1^\top\beta-1|\) & Med.\ ESS & Weak-tangent ESS & TPR & FDR & \(D_{\rm elig}\)\\
\midrule
0   & 9.772 & 0.2557 & 0.6851 & 301.6 & 146.5 & 0.8025 & 0.00268 & 5.49\\
0.1 & 9.785 & 0.2549 & 0.5955 & 302.9 & 146.9 & 0.8063 & 0.00268 & 5.49\\
1   & 9.976 & 0.2518 & 0.3418 & 212.7 & 106.5 & 0.8150 & 0.00125 & 4.31\\
10  &10.346 & 0.2482 & 0.1118 & 302.7 & 160.8 & 0.8200 & 0.00250 & 5.03\\
\bottomrule
\end{tabular}
}%
\end{table}

As discussed in Sections~\ref{sec:introduction} and ~\ref{sec:method}, the same affine information that improves adherence also changes the numerical problem.\ When MYULA is used without geometry-aware preconditioning, a single Euclidean timestep is chosen from the global smoothness constant \(L_f\).\ The effect is already visible at \(\kappa_c=1\) as mixing deterioration:\ median coordinate ESS falls from about \(302\) to \(213\).\ At \(\kappa_c=10\), it becomes pronounced.\ Across the ensemble,
\[
\frac{L_f}{L_\perp}=10.3977
\]
on average, where \(L_\perp\) is the smooth curvature restricted to the tangent space orthogonal to \(\mathbf 1\).\ More explicitly, if \(v_1\) is the unit leading eigenvector of the smooth Hessian and \(u=\mathbf 1/\sqrt p\) is the unit sum direction, then the ensemble mean of \(|v_1^\top u|^2\) is \(0.999366\). Thus the direction of largest curvature is, to numerical accuracy, almost parallel to the sum direction.\ The scalar timestep is therefore governed almost entirely by this single stiff direction.\ Median coordinate ESS falls to \(33.8\), the median ESS in the weakest tangent direction to \(20.1\), and the median minimum decision distance among MAP-eligible variables to \(1.25\).\ In contrast, the sum direction itself has median ESS above \(1100\).\ Therefore, the difficulty is not failing to explore the affine direction;\ it is that curvature in that direction forces the rest of the sampler to evolve much more slowly.

For \(\kappa_c=10\) we consequently use the rank-one preconditioning matrix from Section~\ref{sec:preconditioning},
\[
M_\phi=P_\perp+\phi P_\parallel,
\qquad
\phi=\min\{1,L_\perp/L_f\},
\]
and set the Moreau scale from the tangent curvature \(L_\perp\) rather than from the affine-dominated \(L_f\).\ The mean \(\phi\) is \(0.0962\), and in the transformed geometry the mean smooth-curvature ratio is reduced to \(L_{f,M}/L_\perp=1.022\).\ Using \(M_\phi\) with \(\lambda\) fixed changes the geometry of the Langevin dynamics while leaving the intended smoothed posterior unchanged.\ Here we additionally replace the default \(\lambda=1/L_f\) by \(\lambda=1/L_\perp\).\ Since \(L_\perp<L_f\), this uses a larger Moreau parameter and therefore changes the smoothed approximation itself, not only how that approximation is explored.\ For this reason, we designed this branch primarily as a geometry-aware posterior calculation rather than as a pure preconditioner-only comparison;\ the timestep choice is audited separately in the supplement.

The resulting gain in Monte Carlo reliability is large.\ Relative to the baseline scalar-step \(\kappa_c=10\) calculation, median coordinate ESS increases from \(33.8\) to \(302.7\), the median minimum coordinate ESS from \(9.1\) to \(127.4\), and weak-tangent ESS from \(20.1\) to \(160.8\).\ The median minimum \(D_j\) among MAP-eligible variables increases from \(1.25\) to \(5.03\).\ Taking, in each dataset, the MAP-eligible coordinate whose activation probability lies closest to the posterior-probability threshold, the median standardised separation from that threshold increases from about \(1.25\) to about \(5\) Monte Carlo standard errors.\ Thus the probability-gate decisions are much better resolved relative to Monte Carlo error.\ In contrast, the reported selection summaries move very little:\ mean selected-set size changes from \(6.65\) to \(6.58\), mean TPR from \(0.825\) to \(0.820\), and mean FDR from \(0.0063\) to \(0.0025\).

\begin{figure}[htbp]
\centering
\includegraphics[width=0.74\textwidth]{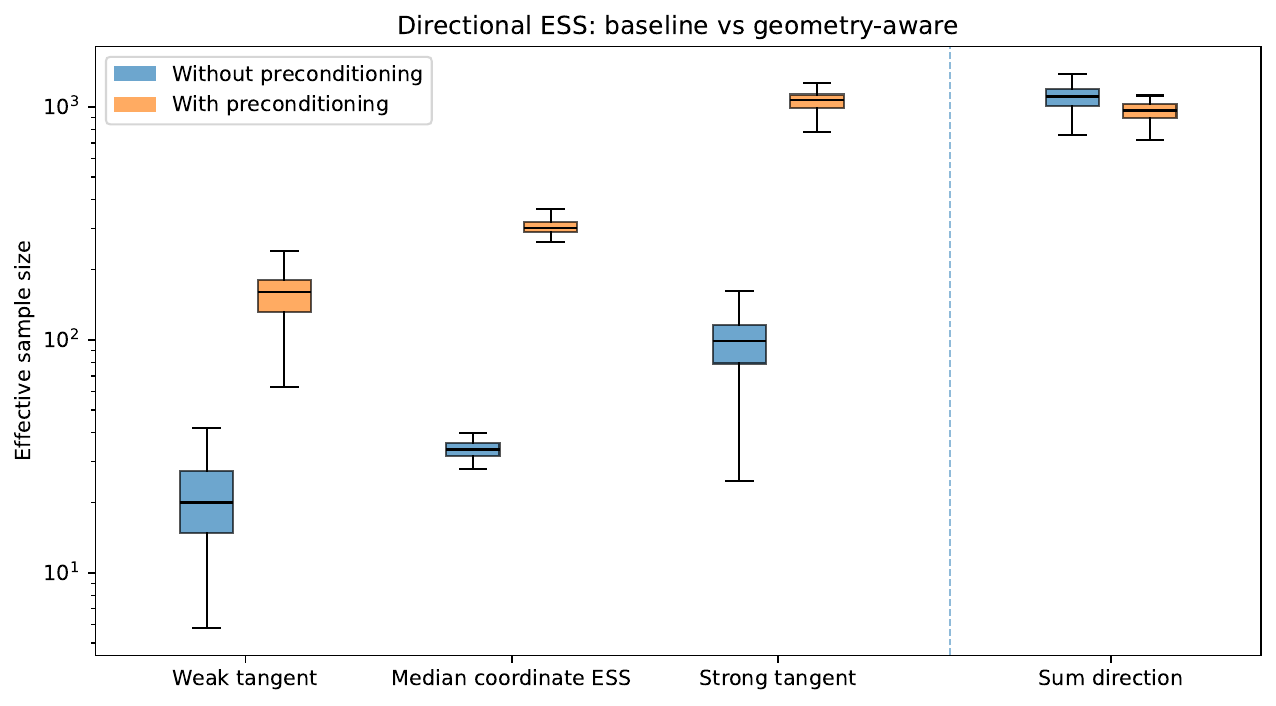}
\caption{Directional effective sample sizes at \(\kappa_c=10\), across \(R=100\) datasets, with and without preconditioning.\ The four summaries are the weak-tangent, median-coordinate, strong-tangent and sum-direction ESS.}
\label{fig:study2-directional-ess}
\end{figure}

Figure~\ref{fig:study2-directional-ess} makes the geometry of the \(\kappa_c=10\) calculation more explicit across the \(R=100\) datasets.\ The weak- and strong-tangent summaries are the ESS values along the least- and most-curved eigendirections of the smooth Hessian restricted to the tangent space, while the median-coordinate ESS gives a robust summary of typical coordinatewise mixing.\ The sum-direction ESS isolates the affine normal direction \(u=\mathbf 1/\sqrt p\).\ Without preconditioning, the sum direction itself is already explored efficiently, whereas mixing deteriorates substantially in the tangent directions and across the original coordinates.\ Preconditioning largely restores that exploration while leaving the sum-direction ESS at the same high order.\ 

\FloatBarrier

Study~2 therefore separates two consequences of softly specifying affine information.\ Strengthening the soft relation genuinely changes the fitted posterior and substantially improves adherence to the relation, with only modest changes in estimation and support recovery.\ At the same time, its rank-one curvature can make a scalar-step posterior calculation unnecessarily inefficient.\ Matching the numerical geometry to that low-rank structure restores decision-level Monte Carlo precision without materially changing the sparse conclusion.

\subsection{Study 3: Diabetes data and external benchmarking}
\label{sec:diabetes}

Study~3 provides an external check on the complete workflow in a real regression problem with correlated predictors.\ We examine calibration, posterior computation, stability of the sparse decision across the prespecified grid, and agreement of the resulting posterior uncertainty with established Bayesian analyses of the same data.

We worked with the diabetes data used by \citet{efron2004}.\ The outcome there is a quantitative measure of disease progression over a year, and the covariates are Age, Sex, Body Mass Index (BMI), average Blood Pressure (BP), and six blood serum measurements (labelled S1-S6).\ There are \(n=442\) observations and \(p=10\) predictors;\ the response is centred and divided by its sample standard deviation, and the predictors are used as supplied.\ The design is full rank, while the Gram matrix has condition number about \(470\).\ The largest absolute pairwise predictor correlation is approximately \(0.897\).\ 
The model is therefore small enough to remain fully interpretable, but contains weak coefficient-attribution directions that make posterior uncertainty nontrivial.\

Accounting for the fitted intercept through the residual degrees of freedom, the classical residual estimator gives
$\widehat\sigma^2=0.493440.$
The extended empirical Bayes calculation gave
$\widehat\theta=0.337289,$
which was then fixed for all downstream calculations;\ the SAPG settings and convergence diagnostics are reported in the supplement.\ We retained the default Moreau choice \(\lambda=1/L_f\).\ In this example, the nonsmoothed FISTA MAP had eight nonzero coordinates, while the corresponding smoothed mode was numerically very close;\ the detailed mode comparison is also reported in the supplement.

The initial scalar-step MYULA run showed slow exploration in the weakest likelihood directions, but the sampler is sufficiently computationally cheap here that extending the run was preferable to introducing an additional preconditioner or retuning \(\lambda\).\ The final calculation therefore retained the same fitted model, Moreau scale and sampler geometry, with a longer simulation horizon.\ Median coordinate ESS was \(1589.6\), the minimum coordinate ESS was \(198.8\), and the weakest-likelihood-direction ESS was \(192.2\).\ The full short-versus-long audit is given in the supplement.

The long-chain 95\% marginal credible intervals exclude zero for Sex, BMI, BP, and S5.\ The final sparse action is stricter than interval zero-exclusion because it also applies the practical magnitude threshold and posterior activation gate from subsection~\ref{subsec: gates}.\ At the reference operating point \((k,\pi_\star)=(2.5,0.75)\), the selected support is
\[
\{\mathrm{BMI},\mathrm{BP},\mathrm{S5}\}.
\]
Thus, the posterior-informed decision reduces the eight-coordinate nonsmoothed MAP support to three coordinates at the reference operating point.\ The full prespecified grid in Table~\ref{tab:diabetesgrid} gives a more informative stability summary:\ BMI and S5 are selected in all nine settings, blood pressure in eight, Sex in two, and every other variable in none.\ 

\begin{table}[H]
\centering
\caption{Diabetes selection grid.\ The reference setting \((k,\pi_\star)=(2.5,0.75)\) is shown in bold.}
\label{tab:diabetesgrid}
\small
\begin{tabular}{c|ccc}
\toprule
 & \(\pi_\star=.50\) & \(\pi_\star=.75\) & \(\pi_\star=.90\)\\
\midrule
\(k=2.0\) & Sex, BMI, BP, S5 & Sex, BMI, BP, S5 & BMI, BP, S5\\
\(k=2.5\) & BMI, BP, S5 & \textbf{BMI, BP, S5} & BMI, BP, S5\\
\(k=3.0\) & BMI, BP, S5 & BMI, BP, S5 & BMI, S5\\
\bottomrule
\end{tabular}
\end{table}
\FloatBarrier

Two coordinates show why the posterior stage adds information beyond the sparse MAP.\ For Sex, the 95\% credible interval lies entirely below zero, but the MAP magnitude 
lies just below the reference practical threshold \(\tau_{\rm post}\).\ 
Its reference activation probability is only \(0.503\), so Sex behaves as a practical-magnitude boundary case, and enters the selected support only at the less stringent \(k=2\) settings with \(\pi_\star\leq0.75\).\ Variable S3 shows the complementary mechanism:\ at \(k=2\) its MAP magnitude exceeds the practical threshold, but its posterior activation probability remains below \(0.5\), so it is selected in none of the nine settings.\ The distinction between MAP eligibility and posterior-supported magnitude is therefore visible in the real-data example itself.

\begin{figure}[!t]
\centering
\includegraphics[width=0.92\textwidth]{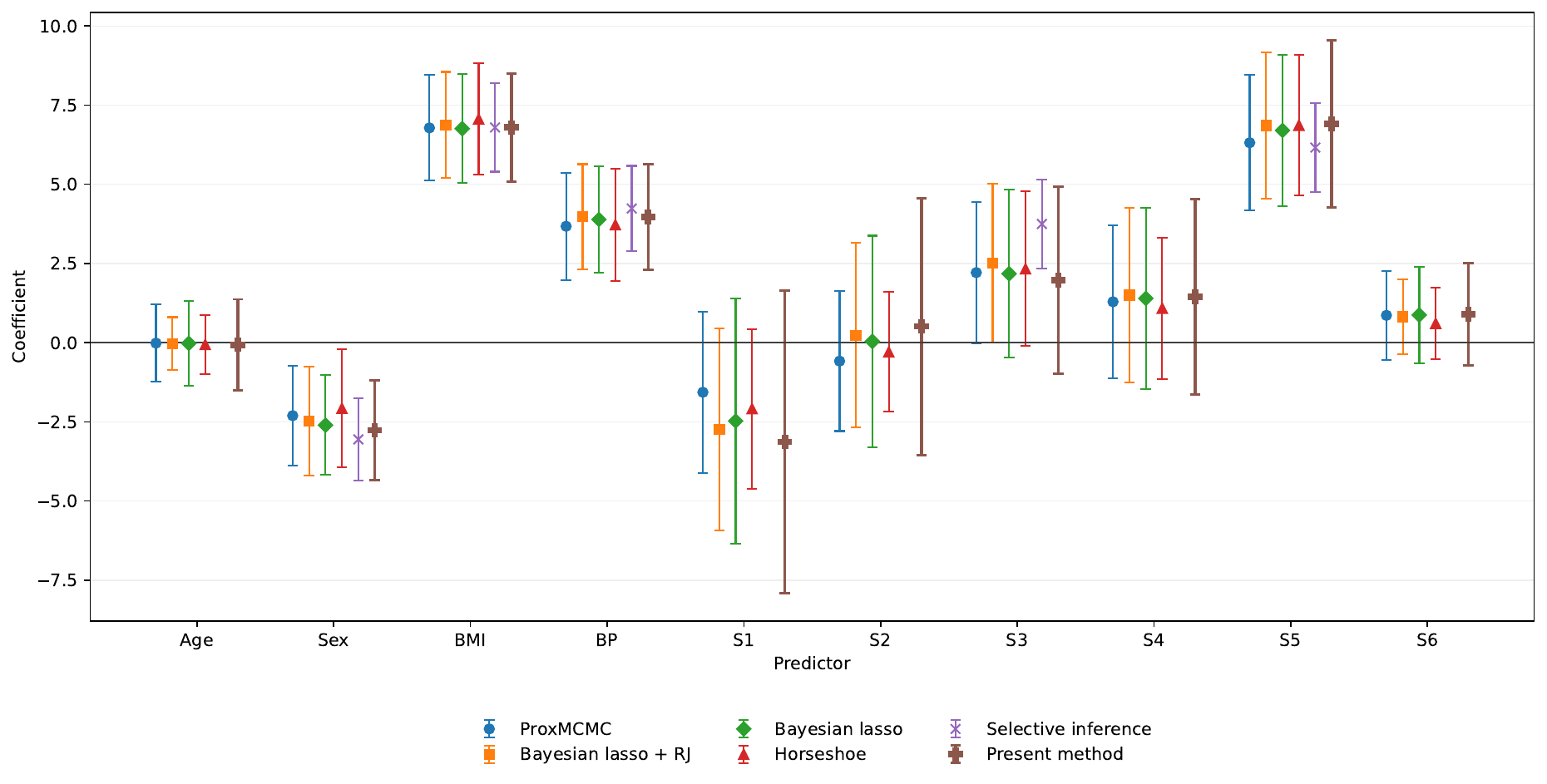}
\caption{95\% coefficient intervals for the diabetes data.\ Benchmark intervals are reconstructed from the public notebook of Zhou et al. (2024);\ the present intervals use the long MYULA run.\ Symbols mark interval midpoints only for visual alignment.}
\label{fig:diabetes-benchmark}
\end{figure}

The same operational data scaling is used in the public diabetes demonstration of \citet{zhou2024}.\ Their accompanying supplementary notebook stores the numerical interval output underlying its published comparison, so we reconstructed that comparison from the authors' recorded numerical values rather than rerunning their mixed Julia/R analysis.\ Figure~\ref{fig:diabetes-benchmark} appends the interval from our long MYULA calculation (EB-Laplace + MYULA analysis) to the recorded 95\% intervals for ProxMCMC, Bayesian lasso with and without reversible-jump selection, and horseshoe intervals, together with the selective-inference intervals reported in the same notebook across Age, Sex, BMI, BP and S1--S6.\ For the reconstructed benchmark methods, centre markers are interval midpoints used only for visual alignment.\ The benchmark reconstruction and all interval endpoints are documented in the supplement.

The Bayesian interval comparisons show the same zero-exclusion pattern.\ ProxMCMC, Bayesian lasso with and without reversible-jump variable selection, the horseshoe, and the present analysis all give exactly the same 95\% Bayesian zero-exclusion pattern:
\[
\{\mathrm{Sex},\mathrm{BMI},\mathrm{BP},\mathrm{S5}\}.
\]
For the strongly identified coefficients BMI, BP and S5, the interval endpoints are also close across the Bayesian procedures.\ The largest differences occur in the correlated serum block, where the present analysis yields wider marginal credible intervals without changing the common zero-exclusion pattern.\ Selective inference additionally reports an interval excluding zero for S3, but those intervals are conditional frequentist objects and the associated lasso support solves a different selection problem.

The interval differences in the correlated serum block are also consistent with the geometry of the regression problem.\ Strong predictor correlation creates weakly identified directions of the likelihood:\ some linear combinations of coefficients are well determined by the data, while movements along other combinations change the fitted response only slightly.\ Posterior mass can therefore extend substantially along these weak directions, and this spread is inherited by the marginal intervals of the individual coefficients even when prediction, or better-identified coefficient combinations, remain comparatively stable.\ Different prior and shrinkage constructions regularise these weak directions differently, so larger differences between Bayesian procedures are most naturally expected precisely in this part of the parameter space;\ in our calculation, extending MYULA greatly increased the effective sample size in the weakest likelihood directions, but the marginal posterior standard deviations did not decrease and, in fact, increased slightly on average.

Study~3 therefore provides two complementary real-data checks.\ The posterior uncertainty from the calibrated proximal workflow is consistent with established Bayesian analyses of the same dataset, while the MAP-plus-posterior rule adds a separate practical-magnitude decision layer.\ Reporting the full decision grid, rather than only its reference operating point, makes that distinction explicit.

\FloatBarrier

\section{Discussion}
\label{sec:discussion}

The main contribution of this paper is a sparse Bayesian regression workflow in which calibration, optimisation, posterior simulation and selection are treated as separate tasks but implemented through a common proximal structure.\ The global shrinkage parameter is calibrated by empirical Bayes, the nonsmoothed MAP remains the reported point estimator, the smoothed posterior supplies uncertainty, and the final rule converts that uncertainty into a sparse decision.\ This separation makes it possible to diagnose where a change in the final support comes from:\ likelihood-scale calibration, shrinkage calibration, the fitted sparse mode, posterior uncertainty, or Monte Carlo error.\ It also retains the numerical advantages of convex regularisation without requiring discrete model-space exploration or a high-dimensional hierarchy of local shrinkage parameters.

The numerical studies show why this separation matters.\ In the $n>p$ regime, scale calibration, MAP recovery and the final decision are all stable, and increasing the posterior run length has little effect on aggregate recovery.\ In the $p>n$ regime, by contrast, weak-signal losses arise primarily from the statistical calibration of the problem rather than from an excess of false discoveries, while the decision-aware Monte Carlo diagnostic identifies individual coefficients whose inclusion remains close to the probability boundary.\ Study~2 exposes a different failure mode:\ strong soft affine information improves adherence to the supplied relation but creates low-rank stiffness that can make a scalar-step Langevin calculation unnecessarily slow.\ Geometry-aware scaling restores effective posterior exploration with little change to the sparse conclusion.\ On the diabetes data, the posterior intervals agree closely with several established Bayesian analyses while the additional magnitude-and-probability rule gives a smaller practical support.

The affine calculations make a further distinction between statistical specification and numerical geometry.\ A hard homogeneous constraint changes the support of the prior and hence the dimension entering the empirical Bayes score.\ A nonhomogeneous affine relation does not admit the same scaling argument, but can be introduced softly without changing the homogeneous prior normalisation.\ That soft formulation preserves the basic SAPG machinery, while its induced curvature can be handled separately at the sampling stage.\ Statistical specification and numerical geometry are therefore related but need not be treated as the same problem.

The framework is modular by construction.\ Observation-scale calibration, empirical Bayes shrinkage, the nonsmoothed MAP, posterior approximation and the final decision can each be changed without redefining the entire procedure.\ The proximal optimisation and sampling components extend directly to other proper closed convex penalties with tractable proximal maps, including elastic net penalties, and Moreau smoothing can be paired with gradient-based samplers other than MYULA.\ When a Gaussian observation-noise interpretation is unavailable, the same architecture can instead use a generalised Bayes learning rate.

There are some limits to the present formulation.\ Plug-in estimates of the observation and shrinkage scales do not propagate hyperparameter uncertainty, and a single global Laplace scale cannot adapt to heterogeneous signals as flexibly as global--local or richer hierarchical priors.\ The final gate is coordinatewise once the joint MAP has supplied the candidate configuration, so strongly correlated problems may benefit from decisions based more directly on joint posterior geometry.\ These limitations do not alter the computational point of the paper:\ a simple convex sparse model can support calibrated shrinkage, a nonsmoothed sparse estimator, posterior uncertainty, affine information and an auditable terminal decision within one reusable proximal architecture.\ The resulting workflow provides a computationally efficient middle ground between reporting a penalised point estimate alone and fitting a substantially richer hierarchical selection model.

\section*{Data and code availability}

The diabetes data are publicly available through the sources associated with \citet{efron2004} and the public notebook accompanying \citet{zhou2024}.\ 

\noindent Code for the reported synthetic experiments, the diabetes analysis and benchmark reconstruction is available at \url{https://github.com/droxanas/ProxEBS_Sparse_Regression}.\ The repository also contains the analysis scripts, fixed random seeds, and the diabetes and benchmark input files needed to reproduce the reported numerical results and figures.

\section*{Supplementary material}

The supplement contains full decision grids, additional SAPG and MYULA diagnostics, MAP-smoothing and preconditioning audits, the complete diabetes benchmark, the hard-constraint calibration check, and the observation-scale/Moreau sensitivity study.\ It is currently accessible from the above repository.


\section*{Statements and declarations}

\textbf{Competing interests.} The author declares no competing interests.\\
\textbf{Funding.} The author received no specific funding for this work.\\

\bibliographystyle{plainnat}
\bibliography{references}

\end{document}